\documentclass[prl,%
 reprint,
 amsmath,amssymb,
 aps
]{revtex4-2}

\usepackage{graphicx} %
\usepackage{dcolumn}  %
\usepackage{bm}       %
\usepackage{hyperref} %

\begin{document}

\title{The Methodology and Implementation of a \\ Real Time Monitoring System for Cryogenic Fridges}

\author{Tabitha Esposito$^1$}
\author{Chloe Greenstein$^2$}
\author{Danielle Speller$^2$}

\affiliation{$^1$University of California, Berkeley, $^2$Johns Hopkins University }

\date{\today}

\begin{abstract}
Inspired by the dilution refrigerator and magnet monitoring system developed for HAYSTAC at Yale University, Speller Lab at Johns Hopkins University developed the Fridge Real Time Monitoring System (FRTMS). The FRTMS accesses logs saved locally by the dilution refrigerator, saves these logs to various backup locations, edits the logs into a format for upload to a MySQL database, and allows the logs to be remotely monitored in near-real time. 
\vspace{\baselineskip}

The python code is publicly accessible through the following GitHub repository:\newline
\url{https://github.com/Speller-Laboratory/JHUFridgeMonitoring-public}.

\end{abstract}

\maketitle

\section{\label{sec:Introduction} Introduction}
Dilution refrigerators are critical in a multitude of research fields and industries, including quantum computing, superconductive materials, and particle physics. The Speller Lab, located at Johns Hopkins University, conducts research requiring temperatures on the order of 10 mK to investigate interactions beyond the standard model of particle physics. The low temperatures provide the environment required to minimize certain backgrounds and achieve the high magnetic fields required for such searches.

The lab’s dilution refrigerator is a complex device which must be continuously monitored through all operational stages in order to prevent or resolve any issues that may impact the fridge infrastructure, experiments, or readout equipment. The fridge (a Bluefors LD400) is a two-stage dilution refrigerator which is first pulse-tube-cooled to 4 Kelvin and then dilution-cooled to a base temperature on the order of 10 mK. The Gas Handling System (GHS) is responsible for circulating the helium, cleaning the mixture, and maintaining vacuum pressures. The cool-down to base temperature takes multiple days, depending on the application. The lab has a 9 Tesla superconducting solenoidal magnet that is supported and thermalized by the 4 K flange of the dilution refrigerator. The extreme low pressures, low temperatures, and high field pose a risk to the equipment and lab safety if conditions vary past operating ranges. Consequently, all pressures and temperatures must be closely monitored and logged during fridge cool-downs. Pressure gauge readings throughout the GHS are recorded by the Bluefors software. The basic thermistor configuration includes readouts for each temperature stage of the cryostat. Two additional temperature sensors are housed in the magnet and there are two empty readout channels for the future installation of additional thermometers.

It is most crucial to monitor the temperature of the superconducting magnet while ramping the field, in order to stop injecting current if conditions approach critical temperature. If the fridge warms past the critical temperature of the material, the magnet risks a quench. 

The necessity to closely monitor the fridge and magnet state at all stages of operation motivated the development of the Fridge Real Time Monitoring System (FRTMS), inspired by Wright Laboratory’s monitoring system for the HAYSTAC experiment at Yale University. The FRTMS allows Speller Lab members to remotely monitor pressures and temperatures of the fridge instead of requiring lab members to come on site multiple times per day. The pressures, temperatures, and any other fridge parameters of interest, can now be viewed remotely at any time via internet access–enabling lab members to come in only when necessary. It also allows for the easy visualization of data from previous cool-downs to reference nominal or unusual behavior.

\section{\label{sec:FRTMSDataFlow} The Fridge Real Time Monitoring System Data Pipeline}

\begin{figure*}
    \centering
    \includegraphics[width=160mm]{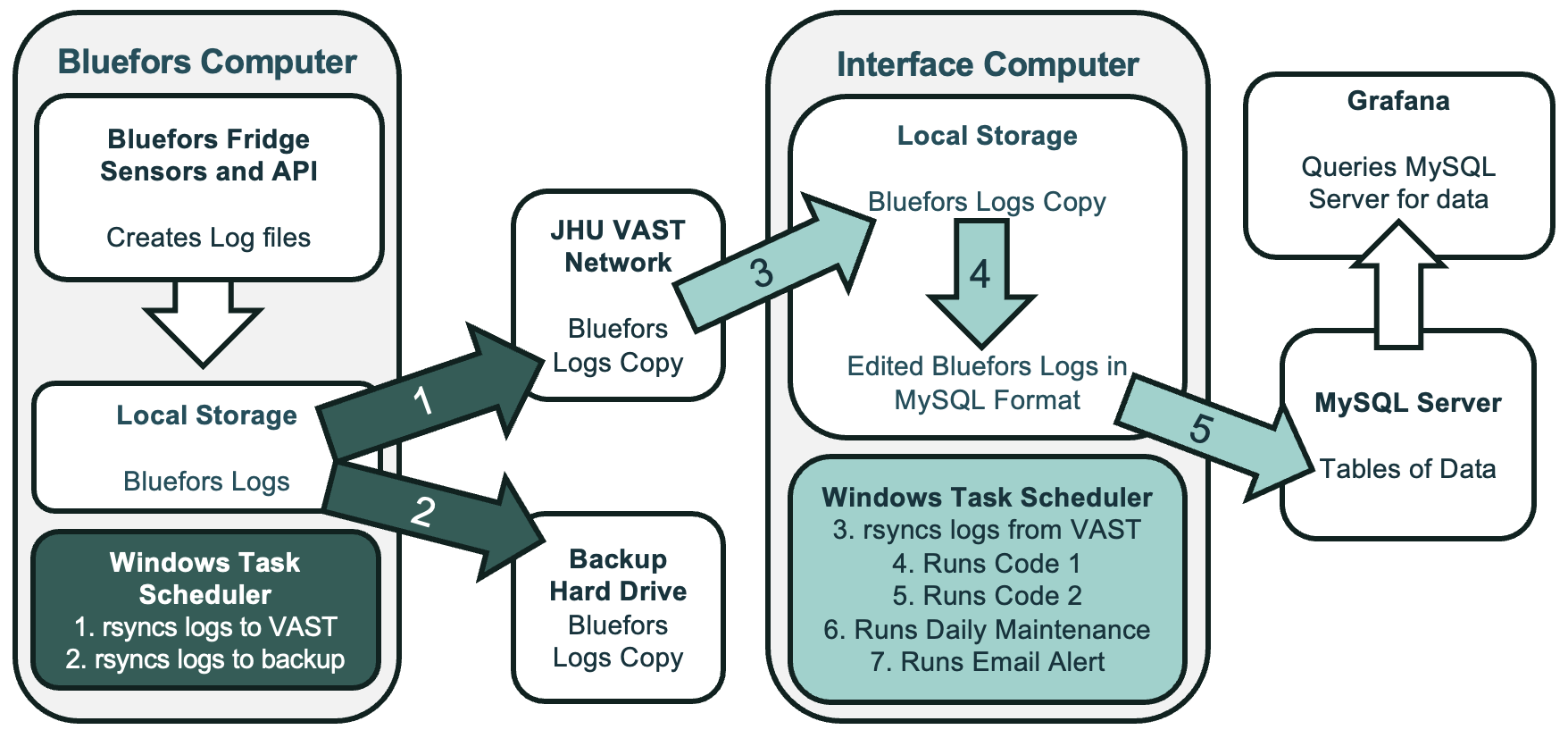}
    \caption{A flowchart describing the hardware and software components of the FRTMS.}
    \label{fig:flowchart}
\end{figure*}

The FRTMS describes the data pipeline required to get logs from the fridge backed up and remotely accessible. The system is composed of two lab computers, a physical hard drive, an external server, a MySQL database hosted on a separate external server, a three-part Python code, synchronizing batch files, Windows Task Scheduler, and Grafana for data visualization, as shown in Figure \ref{fig:flowchart}. All processes are scheduled such that logs from selected fridge parameters are graphed in near-real-time on the Grafana portal. There are active alert systems to contact the Speller Lab members if data uploading lags or values cross a predefined threshold.

The tasks of recording the fridge’s logs, reformatting data, and uploading data are compartmentalized between two computers. The Bluefors Computer (BF) is reserved strictly for controlling the fridge through Bluefors’s Central Control software. Because the BF is used for all fridge activity, data processing is delegated to another computer. The Bluefors Core software generates “.LOG” files in a local folder for all active parameters at user-defined time intervals. Speller Lab records data to the logs every minute. Windows Task Scheduler on BF triggers batch commands that one-way mirror the local folder to a physical, two terabyte, external hard drive and a folder on the VAST server. The batch commands are generated by FreeFileSync, an open source software that builds and tracks batch commands for file comparison and synchronization \cite{freefilesync}. The program keeps track of the last run, the history of batch files created and uploaded to VAST, and syncing errors. VAST is a file system belonging to Johns Hopkins University where Speller Lab has a storage allocation \cite{JHUARCH}.

Windows Task Scheduler triggers batch commands that one-way mirror files on the VAST server to a local folder on a Windows PC known as the ``Interface Computer'' (IC). Windows Task Scheduler then triggers the IC to run three different Python scripts using an Anaconda distribution. 

The first code picks out the specified parameters from the logs and reformats the data into text files, as described in Section \ref{sec:LogDataHandlingCode}. The second code uploads the reformatted files to the MySQL database starting from the last upload and is also described in Section \ref{sec:LogDataHandlingCode}. There is a third email alert code, covered in Section \ref{sec:PythonEmailAlert}, that checks whether log files are successfully pushed to the VAST server and are reformatted continuously without any interruptions. All three codes are continuously called using Task Scheduler on IC.

Generally, all pipeline automation jobs managed by Windows Task Scheduler on BF and the IC are triggered in order every minute with 15-second staggering.

\subsection{Log Data Handling Code}\label{sec:LogDataHandlingCode}
Two main Python scripts process the log files from their raw form to visualization in Grafana. The first two scripts reformat and upload the files continuously using the Python MySQL Connector package to create connections to the database where it submits and pulls queries. This is done in the code using a cursor to execute query statements and store queried data. A third code checks that the system is running and alerts the lab if files are not updated using the Python EmailMessage package.

\vspace{2mm}
\textbf{Code 1: File Reformatting}
\vspace{2mm}\\
The code reformats the data from the raw log files into the correct format to insert into MySQL so that Grafana can successfully query the data as a timeseries. All system specific definitions, such as server usernames, passwords, and filepaths, are defined and referenced through all data processing in an environment script created to compartmentalize user-specific variables without interfering with the code automation. 

For the data to be continuously updated, the code must reopen any files still being updated by the fridge, or which did not finish uploading in the previous iteration. When Bluefors software generates logs, they are stored in subfolders labeled by date. The same goes for the data copy in VAST and in the local IC folder. A table in the MySQL database contains all folder filepaths that have been fully reformatted. Code 1 only opens folders if the folder paths are not in the table to save on code compiling time.

Once in a folder, the files go through a process of sorting for two different case types. First, the files are sorted into two different subcases: current-day files and previous-day files. Current-day files are reopened and written continuously, while previous-day files are not reopened in future iterations once completely written in the current code iteration. The second case type determines whether the file has been completely, partially, or not reformatted since this determines the location from which to begin writing new files.

At this point, the files’ datetimes need to be reformatted, and the correct data needs to be extracted. The raw log files written by the fridge use boolean values to indicate switches and include various parameters that are not of interest to regularly monitor. In addition, sometimes parameters are not recorded, corresponding to empty intervals of timestamps. The code scans each row for the desired data, and if not found for that timestamp, a dummy variable is inserted in lieu of an entry so that the columns of parameters have the same indices in the new file. Some logs ($>$40 columns) are split into multiple text files grouped by similar parameters to alleviate a large number of columns and overcrowded tables. The new files are saved into folders for different runs according to predefined date ranges in the code. The code works the same without sorting runs, but this allows for faster manual access to informative data.

\vspace{2mm}
\textbf{Code 2: Database Upload}
\vspace{2mm}\\
In the second code, the reformatted text files are uploaded into the MySQL database using the Python connector. The code loops through all the reformatted files, only opening them if they have not been already added to a table of filepaths in the database. This table only contains files that have been completely uploaded to the database. The files are sorted again using the same cases as the first code to determine the row within the text file to start uploading to the database. After every upload, the dummy values are replaced in the tables with “null” values. The code is continuously run to process the edited logs and push them to the MySQL database for querying.

\subsection{\label{sec:DatabaseStructure} Database Structure}
The database chosen for this project is MySQL, due to its handing of time series data and compatibility with Grafana, the open source analytics platform available online \cite{grafana}. The database is stored on a JHU server belonging to the Department of Physics and Astronomy. The Speller database has so far used 5.2 GB of data from the database over the past two years of data being stored. The tables created are: (p1-p6) for pressures of the bellows, (temp1-temp8) containing the temperatures for the different channels–including two for the magnet, flowmeter for the flow of the mix, cpa\_status to monitor various parameters of the cpa pump, and turbo\_status to store various parameters of the turbo pump. The tables consist of a first column that is a time series data type and then proceeding columns that are Decimal data types.

Since the data graphed in the main Grafana dashboard only holds up to the last 30 days, another process exists to upload older cool-down data to additional dashboards. In this Python code separate from the first two, edited files are sorted into folders corresponding to run dates, run tables are created in the database for each parameter, and the files are uploaded into their corresponding tables. This code is a modified copy of “Code 2” that uploads the run files, based on dates. The run-sorting code must be run manually and the run dates with their associated numbers are defined in the environment file. First, edited log files saved on IC are moved to their own folder named by run number. The code builds tables in the MySQL database labeled by the run number and uploads the associated data. The associated data is also cleared from its previous filing, effectively converting the logs from general tables containing the last 30 days of data to run specific tables within the database corresponding to a specific range of dates. Each run has a Grafana dashboard which allows for quick referencing. After a new run, a lab member must use the IC to update the environment code with dates, run the notebook, and manually create a Grafana dashboard.

\subsection{\label{sec:Grafana} Grafana}

Grafana is a data visualization application that can be hosted on a server that synchronizes with a data source to continuously query and display a large amount of data. Dashboards are created to contain a group of panels. Each panel displays data from a specific column in a specific table that is queried automatically at user specified time intervals. The panels can show a time series as multiple different types of graphs.

JHU hosts a Grafana server on the Physics and Astronomy department server. The server is accessible outside of the network through a protected login. Parameters currently being monitored are the pressures and temperatures of the different channels in the fridge.

\begin{figure*}
    \centering
    \includegraphics[width=165mm]{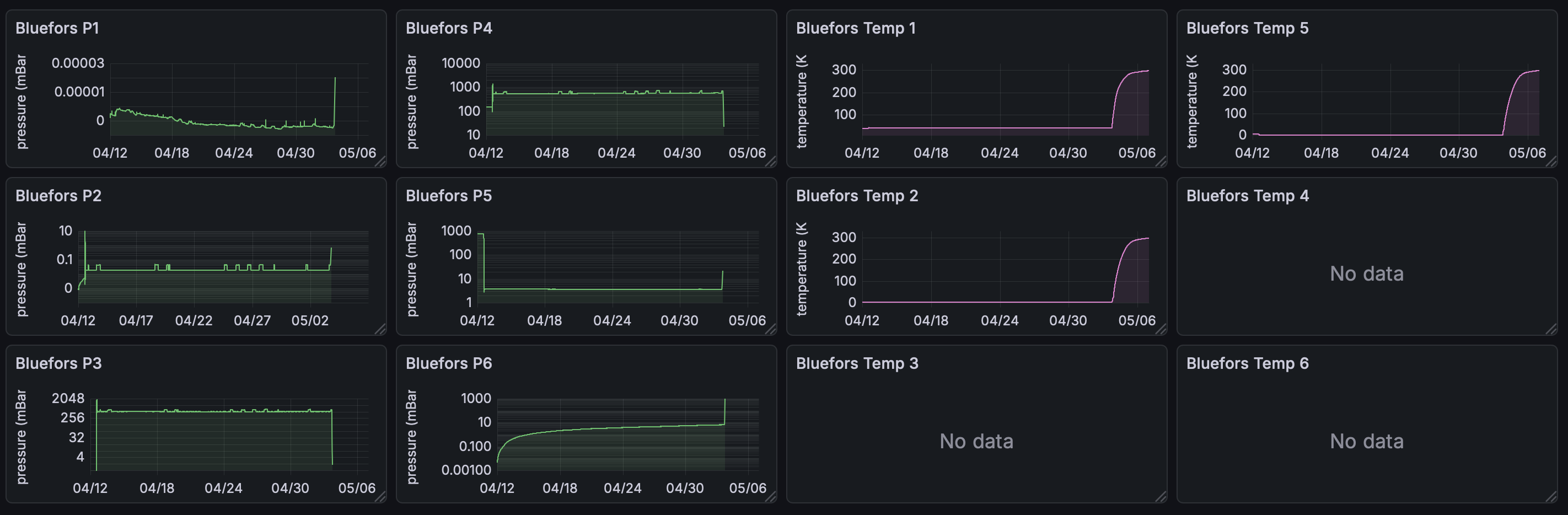}
    \caption{A Grafana Dashboard displaying pressure and temperature data from a single cool-down.}
    \label{fig:GrafanaExample}
\end{figure*}

\section{Alert System}
\subsection{Python Email Alert}\label{sec:PythonEmailAlert}

A Python code is scheduled and run through Task Manager to monitor the log data flow, identify the most recent update time, and determine if there is a lag. The code selects the most recent update in three places: the VAST folder, the edited logs folder on IC, and the MySQL table entries. If the lag time falls between the minimum and maximum time frame, an email is constructed and sent to the lab email detailing the problem.

The third code goes down the log file processing chain, checks the most recent update, decides if there is a lag, and checks if the log files have been uploaded to the VAST server and if the converted text files have been rewritten. This makes sure that the FreeFileSync and the Python code part one is running without any error. Given a maximum and minimum threshold for how long after the current time the files should be uploaded, the code checks how long it has been since the files were last modified in both the VAST server and on IC. If the lag time falls between the minimum and maximum time frame, an email is constructed and sent to lab members detailing the problem. 

\subsection{Grafana Alerting System}\label{sec:Grafana Alerting System}
Grafana has an alerting function where rules may be defined and, depending on the rule definition, messages are pushed out to contact points. Slack notifications were chosen as inspired by HAYSTAC’s procedure and the lab’s consistent use of it as a rapid communication channel. A Slack API was configured with a webhook for Grafana’s contact point \cite{grafana_config_slack}\cite{slack_webhooks}.

There are two different types of alerts that the FRTMS utilizes: lag monitoring, and threshold rules. The lag monitoring backs up the email code to check near-real time logs are continuously visible on Grafana. This rule identifies lags by querying a table for the 50 most recent table entries. The limit is meant to cut down on run time. Only one table needs to be checked to ensure the pipeline is fully transmitting logs. The query then puts a limit to only keep logs older than 20 minutes. If no table entries come from the query, the Grafana assumes normal function. If table entries remain after the filter, an alert is sent. The alert includes the number of seconds since the last upload. This backup alert is necessary to decentralize alerting from the lab. If IC loses power, the data pipeline will pause, and no email alert will be sent. In this case, the lab relies upon the external Grafana and Slack servers.

A threshold rule, on the other hand, checks critical data types and alerts lab members if these values go outside of operating range. These rules all follow the same template in Grafana and are customizable based on the run status. Rules can also be turned on and off depending on stage of operation and allow for changing expected ranges. Each alert rule queries their respective table for the most recent entry. Grafana has a threshold alert condition option to set the alert range. For example, if the pulse tube “water in” is in range 20 $^{\circ}$C to 9000 $^{\circ}$C (to rule out null values), an alert is sent out. In this case, the water has gotten too hot and the pulse tubes should be paused based on lab members' evaluation.

\section{\label{sec:Conclusion} Conclusion}
The FRTMS functions smoothly and has for several cool-downs. The Speller lab is able to monitor fridge cool-downs remotely without the previous necessity of checking every day. The code’s interval of updating every minute is sufficient for data monitoring, so there is no more need to improve the speed of the code. To better help analyze past runs, there are considerations to divide the parameter’s tables in the database by runs. The Python code part two would need to be updated to sort parameters into different databases defined by runs.

\section{Acknowledgements}
Thank you to the HAYSTAC group at Yale University, especially Michael Jewell, who built the monitoring system that inspired this work. We would like to express great thanks to the lab members Beatriz Tapia Ortegui, Rebecca Kowalski, Maryam Esmat, Rouxi Wang and Yongqi Wang for their mentorship and support during the development of the FRTMS. A very special thank you to Dr. Danielle Speller without whom the project would not have been created. She guided the team with expertise on the subject while providing freedom to learn and explore our own solutions. We would also like to thank the Physics Department IT staff, Jessie Warford, Daniel Collura, and Louie Armstrong, for helping secure the servers and troubleshoot. In addition, the lab would like to thank the Advanced Research Computing at Hopkins (ARCH) for hosting the VAST server for the experiment. The FRTMS was created with the support of the Rowland Summer Research Fellowship, the Sloan Research Fellowship in Physics FG-2022-19263, and the Packard Fellowship for Science and Engineering 2022-74683.

\section{Author Contributions}
D.S. conceptualized this work; T.E. and B.T.O. organized the framework; T.E. wrote the log processing code and drafted the FRTMS structure; C.G. automated the process and continued work on the log processing code; T.E. and C.G. both wrote documentation and test code. T.E. and C.G. contributed to this writing equally; D.S. supervised this work and provided writing revisions.

\bibliography{apssamp}

\end{document}